\documentclass[12pt,preprint]{aastex} 

\usepackage{graphicx}
\usepackage[twocolumn]{emulateapj5}

\newcommand{\halpha}{H$\alpha$}
\newcommand{\hbeta}{H$\beta$}

\newcommand{\kms}{km~s$^{-1}$}
\newcommand{\flam}{erg~cm$^{-2}$~s$^{-1}$~\AA$^{-1}$}

\shortauthors{SCHMIDT, SMITH, FOLTZ, HINES}
\shorttitle{SCATTERED BROAD LINE IN A TYPE 2 QSO}

\slugcomment{To appear in the Astrophysical Journal Letters}

\received{2002 July 29}
\revised{2002 Sept. 11}

\begin{document}

\title{An Extraordinary Scattered Broad Emission Line in a Type 2 QSO\altaffilmark{1}}

\altaffiltext{1}{A portion of the results presented here made use of the
Multiple Mirror Telescope Observatory, a facility operated jointly by The
University of Arizona and the Smithsonian Institution.}

\author{Gary D. Schmidt\altaffilmark{2}, Paul S. Smith\altaffilmark{2},
Craig B. Foltz\altaffilmark{3}, and Dean C. Hines\altaffilmark{2}}

\altaffiltext{2}{Steward Observatory, The University of Arizona, Tucson AZ 85721}
\email{gschmidt, psmith, cfoltz, dhines@as.arizona.edu}

\altaffiltext{3}{Multiple Mirror Telescope Observatory, The University of Arizona,
Tucson, AZ 85721}

\begin{abstract}

An infrared-selected, narrow-line QSO has been found to exhibit an
extraordinarily broad \halpha\ emission line in polarized light.  Both the
extreme width (35,000~\kms\ full-width at zero intensity) and 3,000~\kms\
redshift of the line centroid with respect to the systemic velocity suggest
emission in a deep gravitational potential.  An extremely red polarized
continuum and partial scattering of the narrow lines at a position angle
common to the broad-line emission imply extensive obscuration, with few
unimpeded lines of sight to the nucleus.

\end{abstract}

\keywords{quasars: emission lines --- quasars: individual (2MASSI J130005.3+163214) --- galaxies: active --- polarization}

\section{Introduction}

If the Unified Scheme for AGN (e.g., Antonucci 1993) is to be equally
successful for objects of high as well as low luminosity, there should exist a
large number of QSOs whose optical/UV spectra are dominated by narrow emission
lines (Type 2 QSOs or QSO-2s).  Like their Seyfert counterparts, the broad
emission-line regions of such objects are expected to lie within and behind a
nominally toroidal obscuring structure, so that broad lines are most readily
visible in light polarized by scattering off particles above and below the
plane of obscuration. Though a few narrow-line objects with quasar
luminosities have been found with broad permitted lines in infrared and/or
polarized light (esp. {\it IRAS\/} QSOs: Hines et al. 1995, 1999; Young et al.
1996), it is clear that such highly-obscured nuclei are generally not present
in classic lists of objects selected via UV excess.

Far more promising is the AGN catalog being compiled from the Two Micron All
Sky Survey (2MASS; Cutri et al. 2001).  The fundamental selection criterion of
this sample is a color limit ($J-K_S>2$) that weeds out stars while at the same
time favoring extragalactic sources with substantial reddenning toward their
nuclei.  This expectation has received strong support from the discovery of
optical linear polarization in a sizeable fraction of 2MASS QSOs (in some cases
$P\gtrsim10\%$; Smith et al. 2002) as well as in the detection of large
absorbing columns toward the associated nuclear X-ray sources (Wilkes et al.
2002).

The above studies found a surprisingly large number of 2MASS QSOs with the
seemingly paradoxical characteristics of polarization/absorption coupled with
prominent Type 1 (broad emission-line) total-flux spectra, properties that are
not easily explained by the simplest models. Nevertheless, the narrow-line
2MASS AGN provide a number of high-luminosity analogues to Seyfert 2 galaxies
where one might seek the broad, polarized emission lines that signal the
presence of an accretion source powering the nucleus. This paper reports the
first such discovery: a narrow-line QSO with an extraordinarily broad scattered
\halpha\ emission line and a polarized flux spectrum that may yield new
information on the inner structure of AGN.

\section{Observations}

In survey observations, 2MASSI J130005.3+163214 (hereafter 2M130005) showed a
moderately bright and rather red ($B=17.1$, $B-K_S=5.24$) spectrum with a
substantial starlight component and prominent narrow emission lines indicating
a redshift of $z=0.080$.  The 2MASS photometry implies an absolute 2$\mu$m
magnitude $M_{Ks}=-25.84$, well within the range of classical QSOs like the PG
sample (Schmidt \& Green 1983). The coordinates of 2M130005 do not correspond
with a detection in the {\it IRAS} Point Source Catalog but the object is an
11.34 mJy FIRST radio source (Gregg et al. 1996) and it is known to be absorbed
in X-rays (Wilkes et al. 2002).  With a Galactic latitude of 79$^\circ$, the
$R$-band polarization of $P=1.68\%\pm0.14\%$ (Smith et al.  2002) is both
intrinsic and tantalizing. Spectropolarimetry was obtained on three occasions
between 2001 Apr. and 2002 Jul. using the instrument SPOL (Schmidt, Stockman \&
Smith 1992) attached to the 6.5~m MMT and 2.3~m Bok reflector of Steward
Observatory.  Due to the smaller collecting area of the Bok telescope and an
inefficient initial primary mirror coating on the MMT, the earlier data are
dwarfed in quality by the latest epoch from the MMT.  All observations used a
low-resolution grating providing one full order of spectral coverage, typically
$\lambda\lambda4250-8500$.  An entrance slit of 1\farcs1 (MMT; 3\farcs0 for the
Bok telescope) yielded a FWHM resolution of $\sim$17~\AA.  At the alt-azimuth
MMT, the instrument rotator was initially aligned with the parallactic angle
and then set to track to a constant equatorial offset throughout the
observational sequences.  Standard calibration procedures included measurements
of a completely polarized source and of interstellar polarization standard
stars made with identical setups during the same telescope runs.

Data from the individual epochs show qualitatively the same signatures and
polarimetric differences that are entirely attributable to different aperture
sizes, the latter resulting in variable amounts of dilution from the host
galaxy of the QSO.  The data were therefore coadded, weighting according to
the quality of each epoch.  Figure~1 shows the final results in four panels.  A
total of 4.1~hr of integration is represented in the figure, dominated by
$\sim$2~hr with high-reflectance MMT optics.

\section{An Enormous, Scattered H$\alpha$ Emission Line}

The total flux spectrum in Figure 1 is that of an emission-line galaxy, with
prominent [\ion{O}{3}]$\lambda\lambda$4959,5007, [\ion{O}{1}]$\lambda$6300,
[\ion{N}{2}]$\lambda\lambda$6548,6584, [\ion{S}{2}]$\lambda\lambda$6717,6731,
and \hbeta\ superimposed on a predominantly stellar continuum. Narrow \halpha\
can also be distinguished in the [\ion{N}{2}] blend. These lines have FWHM of
$\sim$1000~\kms, i.e., slightly resolved in our data.  Measurements of the line
strengths, including the results of deblending of the
\halpha+[\ion{N}{2}]$\lambda\lambda$6548,6584 complex, yield the following line
ratios:
$\log$([\ion{O}{3}]$\lambda$5007/\hbeta) = 1.03,
$\log$([\ion{N}{2}]$\lambda$6584/\halpha) = 0.14,
$\log$([\ion{S}{2}]$\lambda$6717+6731/\halpha) = $-$0.09, and
$\log$([\ion{O}{1}]$\lambda$6300/\halpha) = $-$0.62.  In the diagnostic diagrams
of Veilleux \& Osterbrock (1987), the gas responsible for the narrow-line
emission is clearly being ionized by an AGN.

The outstanding discovery, however, is an extraordinarily broad emission hump
in polarized flux, $q'\times F_\lambda$, centered at rest wavelength
$\sim$$\lambda$6630.  Interpreting this as scattered broad \halpha\ emission,
its width (18,000~\kms\ FWHM, or $\sim$35,000~\kms\ full-width at zero
intensity) is the largest yet found for scattered permitted lines, and equal to
the broadest observed in {\it any\/} AGN:  cf. the extreme broad-line radio
galaxies 3C~332 (e.g., Eracleous et al. 1997) and 3C~382 (Osterbrock, Koski, \&
Phillips 1976).  The broad \halpha\ line in 2M130005 is redshifted relative to
the narrow lines by $\sim$3000~\kms\ and situated atop an extraordinarly red
continuum.  Polarized, narrow [\ion{O}{3}]$\lambda$5007, [\ion{N}{2}]+\halpha\
are also prominent.  It is important to note that the [\ion{O}{1}] and
[\ion{S}{2}] lines are absent in polarized light, even though the data quality
is certainly adequate to detect the latter if it were of the same relative
strength as the [\ion{O}{3}] and [\ion{N}{2}] lines.  From the line profile
comparison shown as an inset in Figure 1, it appears that
[\ion{N}{2}]$\lambda$6584 is significantly weaker relative to narrow
\halpha\ in the polarized light spectrum than in total light. Finally, despite
all of the structure in polarized flux, the position angle rotates smoothly by
less than 15$^\circ$ across the observed spectrum.

We have attempted to make a correction for the host galaxy's contribution to
the total flux spectrum by scaling a spectrum of the E0 galaxy NGC~3379
(Kennicutt 1992) to the strength of the stellar absorption lines in 2M130005.
Our best attempt at cancelling the G and \ion{Mg}{1} b bands is depicted by the
thin line denoted ``AGN'' in the bottom panel of Figure 1.  The subtraction is
not equally good for all features, but the resulting stellar fraction of 0.8 at
$\lambda$6000 is probably correct to within 10\%. The \ion{Na}{1} D feature
could not be eliminated for any stellar fraction short of unity, and that
choice oversubtracted the blue continuum. The fact that this feature is
marginally apparent in polarized flux suggests that it may be intrinsic to the
scattered light\footnote{See the similar feature in Mrk~231 (e.g., Smith et
al.  1995).}.  It is important to note that the broad \halpha\ component can be
discerned as a shallow hump underlying the narrow lines in the residual AGN
spectrum.  This is true for {\it any\/} reasonable stellar fraction.  A dotted
line indicates the underlying AGN continuum in Figure 1.

Comparison of the peak brightness of broad \halpha\ in polarized and total
light, $\sim$$1\times10^{-17}$~\flam\ {\it vs.\/} $4-5\times10^{-17}$~\flam,
implies a degree of polarization for the scattered light of $P\ge20\%$, with
the inequality resulting from the possibility that some of the broad line in
the total flux spectrum is unscattered.  (This result effectively rules out
dichroic extinction -- the interstellar mechanism -- as the polarizing
agent.)  By contrast, the implied polarization of the starlight-subtracted AGN
continuum is nearly wavelength-independent at $P\approx10\%$ over
$5000-8500$~\AA\ and the continuum-subtracted polarizations of narrow
[\ion{O}{3}]$\lambda$5007 and the [\ion{N}{2}]+\halpha\ complex are
$1.91\%\pm0.20\%$ and $2.58\%\pm0.33\%$, respectively.  Comparing these values
to the measurement for broad \halpha, we infer that $\lesssim$10\% of the
total light in the narrow lines is scattered.  The discrepancy between
broad-line and (galaxy-corrected) continuum polarization is reminiscent of the
situation in some Seyfert galaxies (e.g., Kay 1994; Tran 1995), where dilution
by hot stars may be an explanation (e.g., Gonz\'alez Delgado et al. 1998).

\section{Discussion}

Much has been made of the illumination of particles situated above and below
the toroidal plane by the brilliant nucleus.  For edge-on perspectives, the
scattering of broad-line emission by these particles can provide an indirect
view of the nucleus that we detect as polarized lines like the enormous
\halpha\ feature in 2M130005.  The radius of the inner edge of the torus is
typically inferred by molecular/maser and hot dust emission at $R\sim1$~pc.
Further reprocessing occurs at mid- and far-infrared wavelengths in a possibly
distinct structure of much larger size.  Clouds in the inner torus are thought
to be individually optically thick at most wavelengths (Krolik \& Begelman
1988), but the filling factor may be low, so it is unclear what the overall
vertical optical depth is to visible-light photons in the inner few hundred
pc.  If this region has significant transparency, some of the narrow-line
emission from one lobe of a bipolar narrow line region will pass through the
central plane to be scattered off particles in the other lobe.  With narrow
lines of [\ion{O}{3}], [\ion{N}{2}], and \halpha\ showing a reduced, but
significant, fractional polarization at the same position angle as the
scattered nuclear emission, it is useful to estimate the importance of this
effect.

We write the fraction of the narrow-line emission emitted on one side of the
nucleus that is subsequently scattered by the opposite lobe as
\begin{equation}
{L_{\lambda,{\rm SC}}\over L_{\lambda,{\rm EM}}} \sim {\Omega R e^{-\tau} \over 4\pi}
\end{equation}
where $\Omega$ is the solid angle of the scattering lobe as seen from the
emitting lobe, $\tau$ is the effective vertical optical depth of the nuclear
region, and $R$ is the effective scattering efficiency of the lobe.  We have
assumed isotropic emission for this crude exercise.  Assuming that the
scattering process is also isotropic implies that the observed fluxes are in
the same ratio.  At the end of \S3, we found that the observed ratio was
$\sim$0.1. The quantity $R$ includes variables such as the optical depth in a
lobe and the ratio of scattering to total extinction, and we have difficulty
imagining an overall value $R>0.5$.  For this limit and $\tau$ small, we find
$\Omega\gtrsim\pi$, i.e., a half-opening angle for a scattering lobe
$\beta\gtrsim60^\circ$.

While the opening angle is uncomfortably large in comparison to the observed
angular extents of scattering and [\ion{O}{3}]-emitting lobes in resolved AGN,
an ionization gradient is expected to exist within each cloud, with
high-ionization species like O$^{++}$ predominately confined to the surface
facing the nuclear engine. Hence, a scattering cloud may see brighter narrow
lines than an edge-on observer. An internal ionization structure also suggests
that the backscattering effect would be greater for high-ionization lines than
for low-ionization lines, possibly explaining the absence of
[\ion{O}{1}]$\lambda\lambda$6300,6363 and [\ion{S}{2}]$\lambda\lambda$6717,6731
in the polarized light of 2M130005 (see Figure 1, inset). Finally, because this
explanation places no requirement on the radial location of the scattering
particles, the broad-line emission and continuum could be scattered so close to
the toroidal plane that the light has difficulty emerging over the ``lip'' of
the dusty torus.  The resulting extinction could account for the unusually red
polarized continuum depicted in Figure 1.

A more attractive explanation for the observed polarization spectrum may be
scattering in an extended narrow-line region that contains a global ionization
gradient.  Stratification of the narrow-line region has been established in
radio-loud AGN (Hes, Barthel, \& Fosbury 1993) and Seyfert galaxies (Filippenko
\& Halpern 1984), among others, and can naturally lead to polarization
differences between species (e.g., Hines et al. 1999 and Tran, Cohen \&
Villar-Martin 2000 for the [\ion{O}{3}] and [\ion{O}{2}] lines in IRAS
P09104+4109).  Unfortunately, this mechanism provides no natural explanation
for the extremely red continuum in polarized light.  The
$5000-8000$~\AA\ portion of the latter can be matched to the spectrum of a
typical QSO using a Galactic reddening law with $A_V\sim4$~mag. At shorter
wavelengths the dereddened spectrum turns up too rapidly, suggesting the onset
of a prominent 3000~\AA\ bump. Of course, if dust grains are the scattering
particles, this amount of extinction should be considered a lower limit, since
the spectrum might be bluened by the scattering.  A somewhat smaller continuum
suppression factor is inferred from a comparison of the measured equivalent
width of [\ion{O}{3}]$\lambda$5007, 330~\AA, with the average for low-$z$ QSOs
from Boroson \& Green (1992), $\sim$30~\AA. However, this result is subject to
possible extinction of the narrow-line emission.  Heavy extinction of a
scattered nuclear spectrum is apparently rare for AGN (see Young et al. 1996
for IRAS 2306+0505), and one is left with the impression that the nucleus of
2M130005 is obscured along most sightlines.

As noted above, the broad \halpha{} emission is redshifted by about 3000~\kms\
with respect to the narrow lines and stellar features.  This is substantially
greater than the mean redward shift of \hbeta\ with respect to the systemic
velocity measured in the large samples of McIntosh et al. (1999) and Zamanov et
al. (2002). In fact, the shift observed in 2M130005 is three times larger than
the largest value reported in either study.  This may indicate that scattering
occurs in an outflowing medium (see also Hines et al. 1995).  However, if we
assume that the redshift has a gravitational origin, the ratio of the black
hole mass to the characteristic radius of the broad line-emitting region can be
estimated as ${\rm M}_{\rm BH} / {\rm R}_{\rm BLR} \simeq 10^{6}~{\rm
M}_{\odot} / {\rm AU}$.  It is interesting to note that this is similar to the
result of an independent derivation, ${\rm M}_{\rm BH} / {\rm R}_{\rm BLR}
\simeq 4 \times 10^{5}~{\rm M}_{\odot} / {\rm AU}$, made under the assumption
that the broad-line clouds are in Keplerian orbits about the central black hole
with a maximum orbital velocity equal to the half width at zero intensity of
scattered \halpha\ ($\sim$18,000~\kms).  An estimate for the black hole mass in
2M130005 therefore lacks only a size for the broad-line region from, e.g.,
reverberation mapping (e.g., Kaspi et al. 2000) or the overall accretion
luminosity.  The latter will be possible using wide-band photometry with {\it
SIRTF\/}.  An X-ray spectrum will soon be obtained, and infrared spectroscopy
will be essential for defining the intrinsic spectrum and obscuration of the
nucleus of this new Type 2 QSO.

\acknowledgements{We are grateful to A. Marble for assistance with the
observations, to J. McAfee for skillful operation of the new MMT, and to the
referee A. Laor.  This research was supported by {\it SIRTF\/}/MIPS and
Science Working Group contracts 860785 and 959969 and NSF grants 97-30792 and
98-03072.}


\clearpage

\begin{figure}
\epsscale{0.8}
\plotone{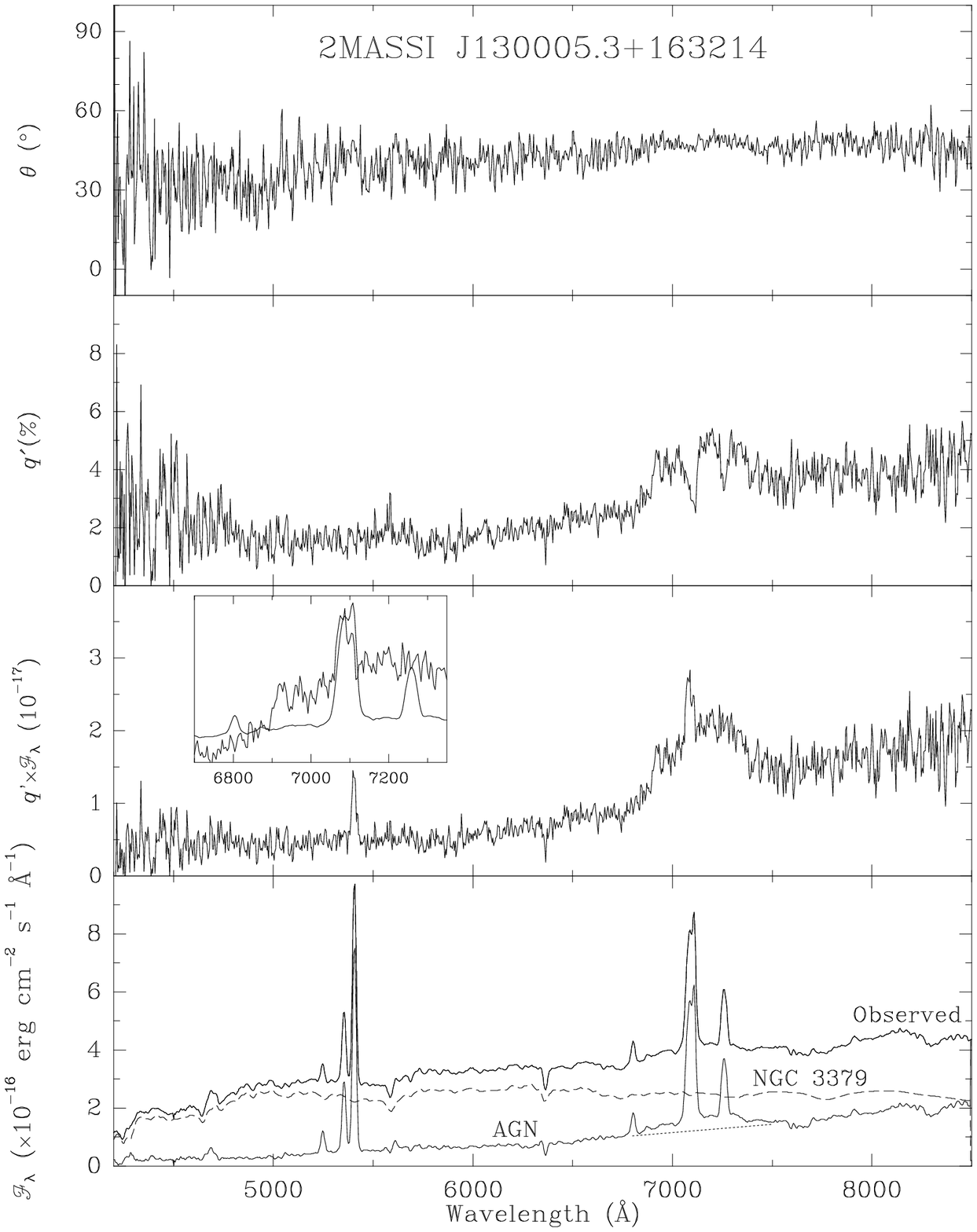}

\figcaption{Spectropolarimetry of 2M130005.  From bottom to top: the observed
total flux spectrum; the Stokes flux $q'\times F_\lambda$ for a frame aligned
with the overall position angle of polarization; the rotated Stokes parameter
$q'(\%)$; and the equatorial position angle $\theta$.  Also shown in the
bottom panel is a scaled spectrum of the elliptical galaxy NGC~3379 used to
subtract the stellar absorption features in 2M130005, as well as the resulting
spectrum of the AGN itself.  The inset compares the narrow \halpha\
spectra in total and polarized flux on arbitrary scales. }

\end{figure}

\end{document}